\documentclass[a4paper, 10pt, twocolumn]{article}

\usepackage[hmargin=1.9cm,vmargin=2.5cm]{geometry}
 
\usepackage[utf8]{inputenc}
\usepackage[T1]{fontenc}
\usepackage[english]{babel}
\usepackage{hyperref}
\usepackage{url}

\usepackage{amsmath}
\usepackage{amsfonts}
\usepackage{amssymb}
\usepackage{mathtools}
\usepackage{lmodern}
\usepackage{upgreek}
\usepackage{ragged2e}
\usepackage{cuted}
\usepackage[version=3]{mhchem}
\usepackage{siunitx}

\usepackage{titling}
\usepackage{titlesec}
\usepackage[affil-it]{authblk}

\pretitle{\begin{center}\large\bfseries}
\posttitle{\par\end{center}}
\preauthor{\begin{center}
\lineskip 0.25em%
\begin{tabular}[t]{c}}
\postauthor{\end{tabular}\par\end{center}}
\predate{\begin{center}\small}
\postdate{\par\end{center}\vskip-1cm}
\date{}

\titleformat{\section}{\normalfont\large\bfseries}{}{0pt}{}
\titlespacing{\section}{0pt}{5pt plus 5pt minus 3pt}{2pt}

\titleformat{\subsection}[runin]{\normalfont\bfseries}{}{0pt}{}[\mbox{ -- }]
\titlespacing{\subsection}{0pt}{4pt}{4pt plus 2pt minus 1pt}

\titleformat{\subsubsection}[runin]{\normalfont\itshape}{}{0pt}{}[: ]
\titlespacing{\subsubsection}{0pt}{3pt minus 3pt}{4pt plus 2pt minus 1pt}

\makeatletter
\renewcommand{\fnum@figure}{\textbf{Figure~\thefigure}}
\makeatother

\usepackage[super,comma]{natbib}
\makeatletter
\def\@biblabel#1{\@ifnotempty{#1}{#1.}}
\makeatother

\usepackage{enumitem}
\usepackage{placeins}
\usepackage{subcaption}
\DeclareCaptionLabelFormat{bold}{(#2)}
\DeclareCaptionJustification{justified}{\justifying}
\usepackage{graphicx}
\usepackage{pdfpages}

\usepackage{varwidth}
\usepackage{xcolor}
\definecolor{mossgreen}{rgb}{0.68, 0.87, 0.68}
\definecolor{mustard}{rgb}{1.0, 0.86, 0.35}
\definecolor{frenchblue}{rgb}{0.0, 0.45, 0.73}
\definecolor{lightskyblue}{rgb}{0.53, 0.81, 0.98}
\definecolor{purpleheart}{rgb}{0.41, 0.21, 0.61}
\definecolor{jasper}{rgb}{0.84, 0.23, 0.24}

\usepackage{tikz}
\usetikzlibrary{calc,fit,shapes,arrows,positioning,shadows,shapes.symbols}

\tikzset{block/.style = {
		rectangle, draw, fill=frenchblue!25, rounded corners, minimum height=3em,inner sep=5pt,%
		drop shadow,
		execute at begin node={\begin{varwidth}{#1}\centering},%
			execute at end node={\end{varwidth}}
	},
	block/.default={9em}
}
\tikzstyle{decision} = [diamond, aspect=1.5, draw, fill=jasper!30, 
drop shadow,
text width=4em, text badly centered, inner sep=0pt,outer sep = -.5pt]
\tikzstyle{cloud} = [draw, ellipse,fill=mustard,
drop shadow,
minimum height=2em]
\tikzstyle{line} = [draw, -stealth, line width=1pt]

\usepackage{cleveref}

\newcommand{\figsublabel}[1]{\protect\phantomsubcaption\label{#1}\protect\bfseries\textsf{\subref*{#1}}}%

\usepackage{xparse} %
\usepackage{xspace}

\ExplSyntaxOn
\prop_new:N \l_pollard_a_prop
\cs_new:Npn \pollard_wrappera:nn #1#2
{%
	\textbf{\subref*{#1}}\nobreakspace\ignorespaces#2\unskip\ignorespacesafterend\hspace{4pt plus 2pt minus 2pt}%
}

\NewDocumentCommand { \addsubcap } {sO{black}m m m m }
{%
	\IfBooleanTF#1
	{\def \fillingArg{}}
	{\def \fillingArg{fill=white}}
	\protect\node[rectangle, text=#2, anchor=north~west,style/.expanded=\fillingArg] at (#4,#5) {\figsublabel{#3}};
	\prop_gput:Nnn \l_pollard_a_prop {#3} {#6}
}
\NewDocumentCommand { \addsubcapNolbl } { m m }
{%
	{\phantomsubcaption\label{#1}}%
	\prop_put:Nnn \l_pollard_a_prop { #1 } { #2 }%
}

\NewDocumentCommand { \processCaptions } { }
{
	\prop_map_function:NN \l_pollard_a_prop \pollard_wrappera:nn
	\prop_gclear_new:N  \l_pollard_a_prop
}
\ExplSyntaxOff

\newcommand{\abs}[1]{\left | #1 \right |}

\begin{document}

\author[1,2,4]{Erwan~Lucas\thanks{erwan.lucas@polytechnique.org}}

\author[1,2]{Su-Peng Yu}

\author[1]{Travis C. Briles}

\author[1,3]{David R. Carlson}

\author[1,2]{Scott B. Papp\thanks{scott.papp@nist.gov}}

\affil[1]{Time and Frequency Division, National Institute of Standards and Technology, Boulder, CO USA}
\affil[2]{Department of Physics, University of Colorado, Boulder, CO 80309, USA}
\affil[3]{Octave Photonics, Louisville, CO 80027, USA}
\affil[4]{Current affiliation: Laboratoire ICB, UMR CNRS 6303, 21078 Dijon, France} 
\title{Tailoring microcombs with inverse-designed, meta-dispersion microresonators}

\maketitle
\thispagestyle{empty}
\noindent\textbf{\boldmath
	Nonlinear-wave mixing in optical microresonators offers new perspectives to generate compact optical-frequency microcombs, which enable an ever-growing number of applications.
	Microcombs exhibit a spectral profile that is primarily determined by their microresonator's dispersion; an example is the $ \operatorname{sech}^2 $ spectrum of dissipative Kerr solitons under anomalous group-velocity dispersion.
	Here, we introduce an inverse-design approach to spectrally shape microcombs, by optimizing an arbitrary meta-dispersion in a resonator.
	By incorporating the system's governing equation into a genetic algorithm, we are able to efficiently identify a dispersion profile that produces a microcomb closely matching a user-defined target spectrum, such as spectrally-flat combs or near-Gaussian pulses.
	We show a concrete implementation of these intricate optimized dispersion profiles, using selective bidirectional-mode hybridization in photonic-crystal resonators.
	Moreover, we fabricate and explore several microcomb generators with such flexible `meta' dispersion control.
	Their dispersion is not only controlled by the waveguide composing the resonator, but also by a corrugation inside the resonator, which geometrically controls the spectral distribution of the bidirectional coupling in the resonator.
	This approach provides programmable mode-by-mode frequency splitting and thus greatly increases the design space for controlling the nonlinear dynamics of optical states such as Kerr solitons.
}
\begin{figure*}[t!]
	\centering
	\begin{tikzpicture}[inner sep=0]
		\node[above right] (img) at (0,0) {\includegraphics[width=\textwidth]{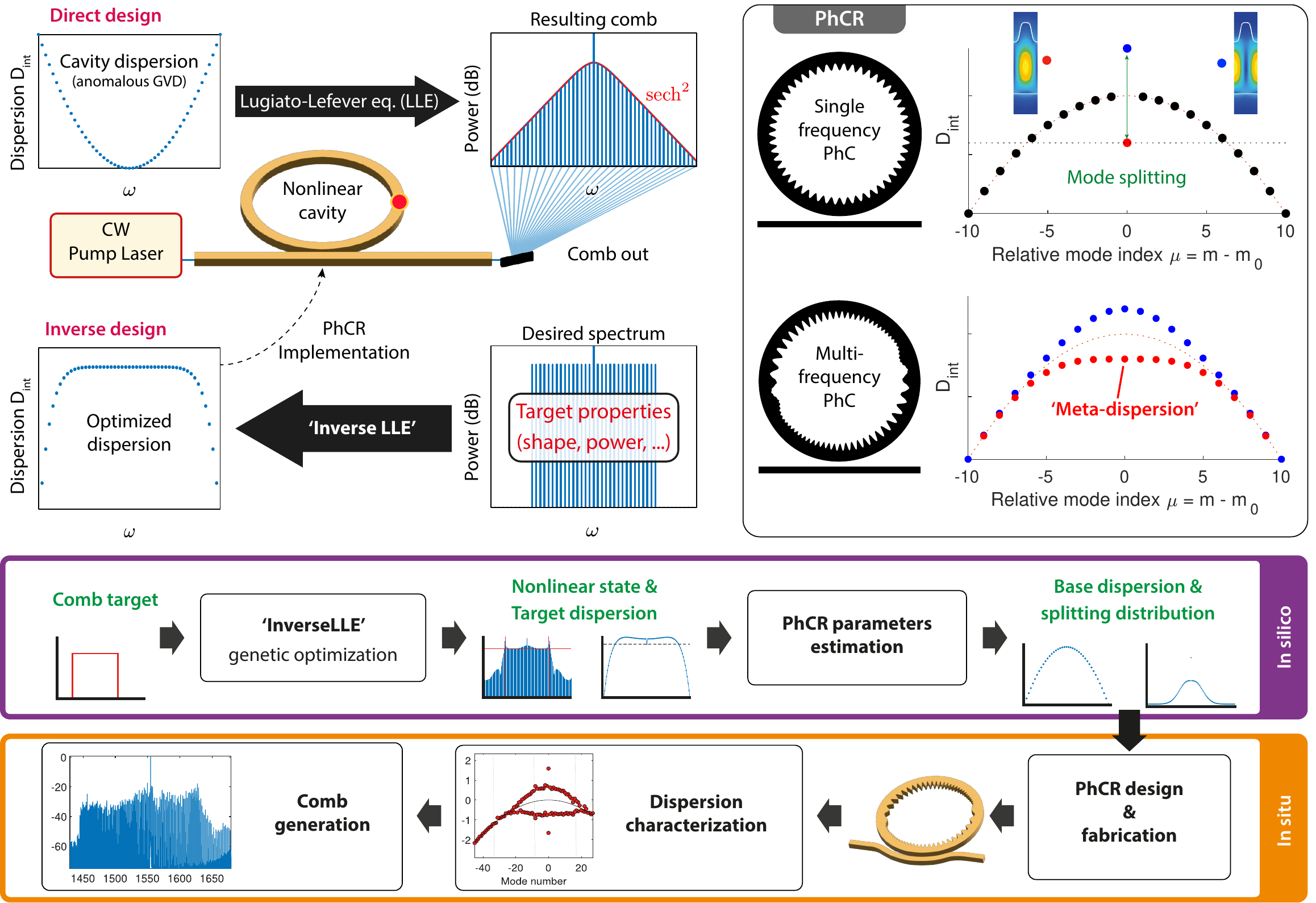}};
		\begin{scope}[
				x={($0.01*(img.south east)$)},
				y={($0.01*(img.north west)$)}
			]
			\addsubcap{fig:concept:DirectDesign}{0}{99.5}{
				Microcomb generation in an optical microresonator.
				The resonator dispersion determines the comb spectral envelope, following the Lugiato-Lefever Equation (LLE).
				Applying this nonlinear model, knowing the resonator parameters, can be viewed as direct design.
			}
			\addsubcap{fig:concept:InvDesign}{0}{64.5}{
				The concept of inverse design consists in the reverse operation, i.e. assignment of a desired comb envelope and determination of the optimal microresonator dispersion for which the resulting comb shape is closest to the target.
				This `inverse LLE' is accomplished here with a genetic algorithm (GA).
				The optimized dispersion profiles are implemented using photonic crystal ring resonators (PhCR).
			}
			\addsubcap{fig:concept:PhCR}{53.5}{100}{
				A PhCR is a micro-ring resonator where the inner ring waveguide wall is patterned with a sub-wavelength modulation ($ \lambda/2 $).
				This leads to a targeted and tunable single-mode splitting, with the formation of a red- and blue-shifted pair of modes (resp. shifted to lower and higher frequency).
				The PhCR approach can be generalized to split multiple modes simultaneously, by superposing several modulation patterns (bottom panel).
				The combined effects of waveguide dispersion and the mode splitting  create a composite dispersion, or meta-dispersion on the  blue- and red-shifted branches.
			}
			\addsubcap{fig:concept:pipeline}{0}{42}{
				Our procedure for tailoring microcombs consists in several steps.
				A desired comb shape is first defined.
				A GA calculates the optimized dispersion profile, which is then converted into meta-dispersion parameters, determining the geometry of the PhCR.
				The devices are nano-fabricated and characterized, before pumping the suitable resonators.
			}
		\end{scope}
	\end{tikzpicture}
	\caption{%
		\textbf{Concept of microcomb inverse-design via meta-dispersion engineering}
		\processCaptions
	}
	\label{fig:concept}
\end{figure*}

Microcombs --- optical-frequency combs generated in driven Kerr resonators~\cite{Kippenberg2018} --- are versatile light sources that offer unique properties for applications and integration of frequency-comb systems on a chip.
Through experimentation and advances in the fabrication of integrated resonators, microcombs have progressed from a theoretical framework for Kerr-nonlinear optics~\cite{Lugiato1987,Coen2013} to the generation of octave spanning combs for accurate and precise optical metrology~\cite{Spencer2018}.
The large mode spacing and spectral width of microcombs have already found applications in parallel coherent communication~\cite{Marin-Palomo2017} distance measurement~\cite{Trocha2018,Suh2018,Riemensberger2019}, and  phase stabilization~\cite{Drake2019a} for photonic-integrated frequency synthesis~\cite{Spencer2018}.

Microcomb generation often involves the formation of dissipative Kerr solitons (DKS)~\cite{Leo2010,Herr2013}, which exist by balancing losses through nonlinear gain, anomalous group velocity dispersion (GVD), and pump laser detuning, through nonlinear phase shifts.
Dominated by second-order dispersion, the resulting DKS pulses feature a squared secant hyperbolic profile in time and frequency domain; see \cref{fig:concept:DirectDesign}.
Owing to this equilibrium with the Kerr-nonlinearity, the dispersion of the resonator -- or more generally mode-by-mode frequency mismatch from the resonator's free-spectral range (FSR) -- primarily determines the pulse shape and spectrum of nonlinear states it supports.

Tailoring the comb spectral profile is desired in applications like telecommunications where high power per mode and spectral flatness improve the performance and efficiency of data links.
However, there is currently no direct relationship connecting the desired microcomb spectrum with the required dispersion, material properties, or physical geometry of the resonator.
Additionally, achieving precise dispersion control in resonators has historically been a significant challenge, lacking a demonstrated technique for imparting arbitrary dispersion properties.
Nevertheless, recent advancements in numerical computation and machine learning have introduced new capabilities and tools for system design and optimization~\cite{Lorenc2007,Sigmund2011,Jensen2011,Genty2020}.
Integrated photonics, on the other hand, not only offers potential for miniaturization of large-scale systems but also enables greater control over guided modes of light, particularly in photonic crystals~\cite{Joannopoulos1997}, or topology-optimized elements~\cite{Molesky2018,YuGeneticallyOptimized2017} which paved the way for highly customizable inverse-designed microresonators~\cite{AhnPhotonicInverse2022} 

Here, we propose and implement an inverse-design process to shape the spectral envelope of microcombs.
Our approach numerically optimizes an arbitrary dispersion profile to tailor the associated nonlinear state of the resonator toward a target pulse shape and spectrum.
The resulting complex dispersion relation is then implemented indirectly by means of Photonic Crystal Resonators (PhCRs), where the resonance frequencies can be controlled via multiple selective mode splitting~\cite{Yu2020,Lu2020,MoilleFourierSynthesis2023}  to create a meta-dispersion (\cref{fig:concept:PhCR}).
We demonstrate a full iteration of the design process and first experimental evidence of microcomb shaping.
The design steps followed in our work are summarised in \cref{fig:concept:pipeline}.
First, we numerically solve the governing equation of the resonator, and use a genetic algorithm (GA) to iteratively optimize the dispersion and pumping parameters to create a desired microcomb shape.
While a similar GA-based method was concurrently developed~\cite{ZhangInverseDesign2022}, our approach notably introduces an analytical model based on the target comb's Kerr shift to efficiently generate an initial dispersion.
Second, we determine the PhCR topology needed to achieve the optimized dispersion.
Finally, we demonstrate realizations of meta-dispersion in PhCR microresonators and initial observations of comb generation in these devices.

\section{Results}
\subsection{Inverse-design dispersion optimization}
We first describe our precise formalism and notation.
The Lugiato--Lefever equation~\cite{Lugiato1987} (LLE) accurately models the driven nonlinear resonator and can be written as a set of coupled mode equations in the spectral domain:
\begin{equation}\label{eq:LLEeq}
	\begin{aligned}
		\frac{\partial \tilde{A}_\mu}{\partial t} & = - \left[
		\dfrac{\kappa}{2}
		+ i \, \left( \delta\omega - D_{\rm int}(\mu) \right) \right] \tilde{A}_\mu                                   \\
		                                          & + i \, g_0 \,  \mathcal{F}\left[\left|A\right|^2  A  \right]_\mu
		+ 	 \sqrt{ \kappa_{\rm ex} \, \dfrac{P_{\rm in}}{\hbar\omega_0} }   \, \delta_{\mu = 0}
	\end{aligned}
\end{equation}
where $ \mu $ is the azimuthal mode number relative to the pumped mode, $ \abs{ \tilde{A}_\mu(t)}^2 $ is the number of photons in mode $ \mu $ as a function of the `slow' time $ t $, $ P_{\rm in} $ and $ \delta\omega = \omega_0 - \omega_{\rm pump}  $ are respectively the pump power and detuning terms~\cite{Herr2013} (the detuning is positive if the laser frequency is lower than the resonance frequency),
$ \kappa $ is the resonator FWHM linewidth, and $ \kappa_{\rm ex} $ is the coupling rate.
$ \mathcal{F}\left[ \cdot \right]_\mu $ represents the Fourier series operator and $ g_0 $ the per photon Kerr shift of the resonator modes.
$ P_{\rm th} = \hbar\omega_0\kappa^3/8g_0\kappa_{\rm ex} $ is the threshold pump power for initial four-wave mixing in resonators and is useful to rescale the power quantities.
The operator $ D_{\rm int}(\mu) = (\omega_\mu - \omega_0) - D_1 \, \mu $ represents the resonator dispersion~\cite{Brasch2015} and measures the deviation of the resonance frequencies from an equidistant FSR $ D_1 $.

\subsubsection{Optimization problem}
The dispersion plays a key role in determining the temporal and spectral shape of localized structures emerging from nonlinear effects in a resonator.
One problem of interest is to find the dispersion $D_{\rm int}$ of the resonator that yields a comb with desired properties, and is a steady-state solution of the LLE; see \cref{fig:concept:InvDesign}.

To simplify the problem, we describe the dispersion using a set of polynomial coefficients $ D_k $ such that
$
	D_{\rm int} (\mu) = \sum_{k=2}^{8} \frac{D_k}{k!} \mu^k + \frac{\gamma_0}{2} \, \delta_{\mu = 0} \;.
$
The extra degree of freedom $\gamma_0$ corresponds to a discrete shift in the frequency of the pump mode, which is known to favour the formation of pulses, notably in the normal dispersion regime~\cite{Lobanov2015,Yu2020,Yu2021}.
An additional constraint limits the pump power within a specified budget of $P_{\rm max}$.
Our objective is to determine the values of $D_k$, $\gamma_0$, $\delta\omega$, and $P_{\rm in}$ that result in a stationary comb profile that best satisfies our objective, as quantified by the error function $\mathcal{E}(\tilde{A}\mu)$ (see the Methods section for details).
To solve this optimization problem, we use a genetic evolutionary algorithm~\cite{HollandAdaptationNatural1992,Kumar2010,ZhangInverseDesign2022}.

\subsubsection{Initialization}
\def \ATargf{\tilde{A}^{\rm target}_\mu}
\def \ATargt{A^{\rm target}}
It is useful to leverage prior knowledge to initialize the problem as close as possible to a potential solution.
An initial dispersion profile is first derived based on a simple heuristic, inspired by the essential principle of the soliton balance of dispersion by the Kerr shift.
Thus, in the frequency domain, the dispersion $ D_{\rm int}^{\rm initial}(\mu) $ is initialized as the opposite of the Kerr spectral shift $ \delta^{\rm Kerr}_\mu $ of the target comb state $ \ATargf $, according to~\cite{Yu2020}
\begin{equation}\label{eq:KerrShift}
	D_{\rm int}^{\rm init}(\mu) = -  \delta^{\rm Kerr}_\mu =%
	- \Re\left(
		\dfrac{ \mathcal{F} \left[ \abs{\ATargt}^2 \, \ATargt \right]_\mu}%
		{ \ATargf } \right)
\end{equation}
Hence this relation is fitted with a polynomial to obtain the initial dispersion coefficients.
The pump laser parameters ($ P_{\rm in} $, $ \delta\omega $) are also initialized, below $ P_{\rm max} $ for the former, and based on the maximum single-mode Kerr-shift for the latter~\cite{Coen2013a}.
A population of dispersion candidates, with a typical size of $ N_{\rm pop} =208 $, is then generated from random variations around these initial parameters.

\begin{figure*}[!htb]
	\centering
	\begin{tikzpicture}
\begin{scope}[font=\footnotesize\sffamily, node distance = 2em, auto, line width=.5pt, every shadow/.style={shadow xshift=1mm, shadow yshift=-1mm,opacity=0.25}]
\node [block=4cm] (init) {\textbf{Initialize population}:\newline $N_{\rm pop}$ individuals\\ \{dispersion \& driving parameters\}};

\node [block=3.5cm, right=of init.north east, anchor=north west] (identify) {Compute \textbf{LLE steady state comb} for each individual};

\node [block, right=of identify] (evaluate) {\textbf{Evaluate} each comb's fitness};

\node [cloud, above=2em of identify] (target) {\bfseries Target comb shape};

\node [block=4cm, below=1.5em of evaluate] (update) {Individuals \textbf{selection}  \&  \textbf{mixing} (crossovers) + \textbf{mutations}};

\node [decision, right=of evaluate] (decide) {Fitness \textbf{criterion} fullfiled?};

\node [block, right=of decide] (stop) {Stop};

\node [block=2cm, below=1.5em of stop,fill=mossgreen] (outp) {\bfseries Dispersion\\profile};

\path [line] (init.east|-identify) -- (identify);
\path [line] (identify) -- (evaluate);
\path [line] (evaluate) -- (decide);
\path [line] (decide) |- node [near start] {No} (update);
\path [line] (update) -| node[below] {New generation} (identify);
\path [line] (decide) -- node[near start] {Yes} (stop);
\path [line] (stop) -- (outp);

\path [line, draw=purpleheart, dashed, text width=2.5cm, align=center] (target) -| node[pos=.7,color=purpleheart,anchor=east,inner sep=0pt, outer sep= 2.5pt]{\bfseries Kerr shift heuristic $D_{\rm int}^{\rm init} = -  \delta^{\rm Kerr}_\mu$}  (init.60);

\path [line, draw=purpleheart, dashed] (target) -| node[color=purpleheart] {Fitness metric} (evaluate);

\node[draw,thick,rounded corners, inner sep=5pt, fit={(current bounding box.south west) (current bounding box.north east)}] (box) {};
\node[anchor=north east] (inverselle) at (box.north east) {\large\bfseries InverseLLE};

\end{scope} 		
		\node[anchor=north west] (lb1) at (box.north west) {\figsublabel{fig:inverselle:algo}};

		\node[inner sep=0pt, below=.5em of current bounding box.south east, anchor=north east] (img)  {\includegraphics[width=\textwidth]{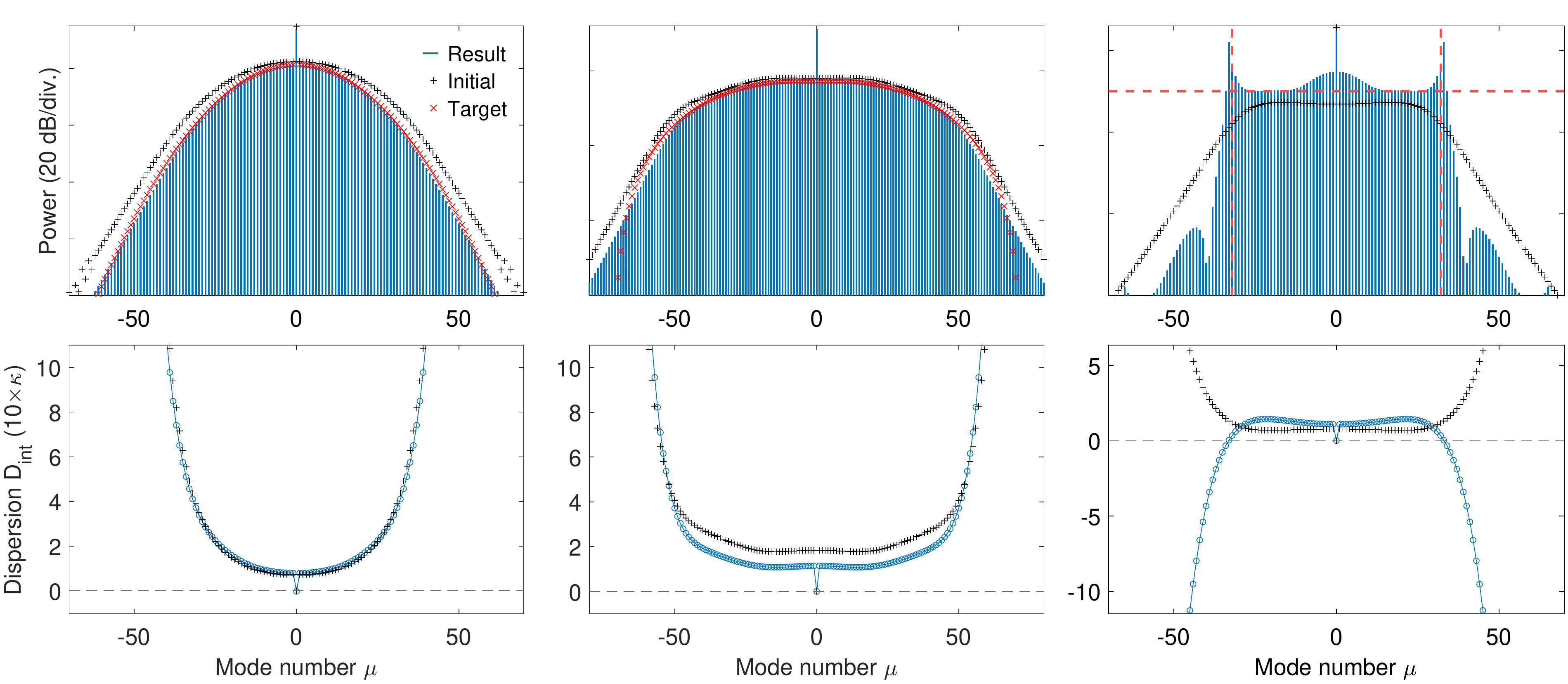}};
		
		\begin{scope}[shift={(img.south west)},
			x={($0.01*(img.south east)$)},
			y={($0.01*(img.north west)$)},
			inner sep=0
			]
			\addsubcap{fig:inverselle:Case1C}{5}{94}{
				Comb spectra of a targeted Gaussian comb envelope (red), initial comb profile computed from the Kerr shift initialization (black dashed), and the evolutionary algorithm result (blue).  The driving parameters $ \{2 \delta\omega / \kappa, \, P_{\rm in} / P_{\rm th}\} $ are initially $ \{4.4, 13\} $ and $ \{5, 8.6\} $ after optimization.
			}
			\addsubcap{fig:inverselle:Case1D}{5}{47}{
				Computed the Kerr shift-based initialization (black cross), and the integrated dispersion profile after  evolutionary optimization (blue).
			}
			\addsubcap{fig:inverselle:Case2C}{38}{94}{
				Set of comb spectra for a targeted raised cosine shape ($ \beta =  0.8$, see methods). Initial driving: $ \{4.4, 13\} $, and optimized set: $ \{6.5, 11.7\} $.
			}
			\addsubcap{fig:inverselle:Case2D}{38}{47}{
				Corresponding dispersion for the spectra shown in \subref*{fig:inverselle:Case2C}.
			}
			\addsubcap{fig:inverselle:Case3C}{71.2}{94}{
				Optimization with the objective of reaching a minimum power per line requirement over a given bandwidth. Initial driving: $ \{3.5,7\} $, and optimized set: $ \{1.2, 7\} $.
			}
			\addsubcap{fig:inverselle:Case3D}{71.2}{47}{
				Corresponding dispersion for the spectra shown in \subref*{fig:inverselle:Case3C}.
				While the initial dispersion is in the anomalous regime, the optimized dispersion is mainly normal.
			}
		\end{scope}
	\end{tikzpicture}
	\caption{%
		\textbf{Dispersion optimization for comb shaping}
		\textbf{\subref*{fig:inverselle:algo}} Flowchart of the evolutionary algorithm used to optimize the dispersion and driving parameters to match a given comb spectral profile.
		\processCaptions
	}
	\label{fig:inverselle}
\end{figure*}
\begin{figure*}[!htb]
	\centering
	\begin{tikzpicture}
		\node[below=1em of current bounding box,inner sep=0] (img)  {\includegraphics[width=\textwidth]{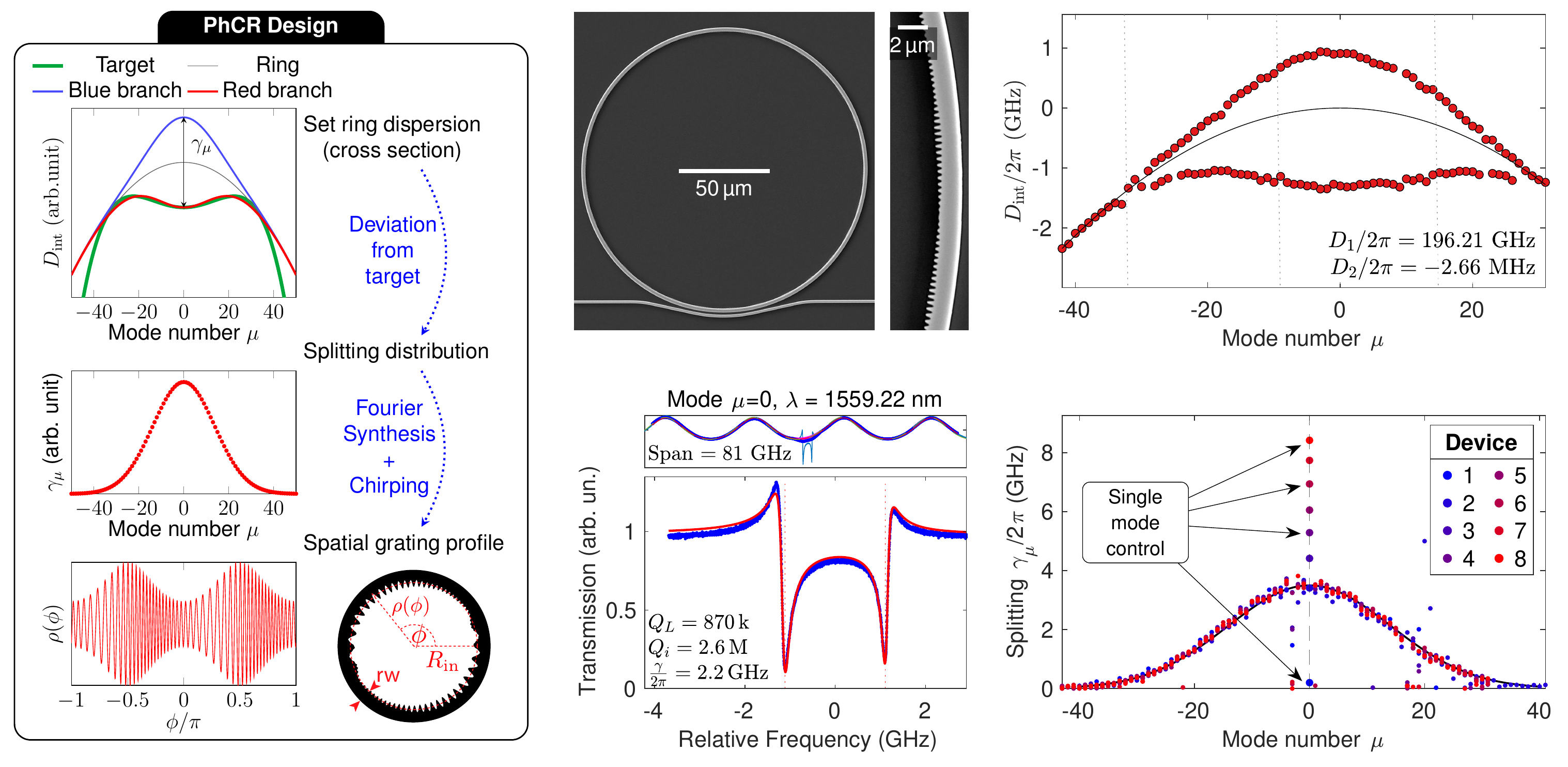}};

		\begin{scope}[shift={(img.south west)},
				x={($0.01*(img.south east)$)},
				y={($0.01*(img.north west)$)},
				inner sep=0, outer sep=0
			]
			\addsubcap{fig:MetaDisp:PhcDesign}{1}{99}{
				Design principle of the meta-dispersion PhCRs.
				First, the resonator dispersion is chosen to match the dominant sign of the targeted dispersion returned by the GA, which sets the waveguide cross-section.
				The spectral distribution of mode splitting is then set equal to twice the deviation between the target dispersion and that of the waveguide near the central modes.
				Finally, the required grating profile is generated by Fourier synthesis.
				The various components of the grating are superposed with a chirp in order to distribute the grating modulation more evenly along the ring.
			}
			\addsubcap*[white]{fig:MetaDisp:SEM}{37}{97}{
				SEM image of a meta-dispersion PhCR. The inset on the right highlights a section of the chirped corrugation.
			}
			\addsubcap{fig:MetaDisp:DintGauss}{69}{97}{
				Measured integrated dispersion of a PhCR (rw=\SI{2.3}{\micro\meter}, th=\SI{570}{\nano\meter}).
				The designed splitting profile follows a Gaussian curve.
			}
			\addsubcap{fig:MetaDisp:pumpResFit}{36}{46}{
				Fitting of the split resonance of the mode $ \mu=0 $.
				The asymmetric lineshape \cite{Fan2002} originates from interference with the chip Fabry Perot (facets reflections) fitted on the top panel.
				A mode splitting model \cite{Gorodetsky2000} is then fitted to the PhCR resonance to extract the loaded and intrinsic quality factors ($ Q_L, Q_i $) and splitting parameters.
			}
			\addsubcap{fig:MetaDisp:gammaPumpTune}{69}{44}{
				Distribution of mode splitting retrieved from fitting the resonances of several PhCR devices.
				The splitting distributions are designed with the same Gaussian profile, where only the splitting of the zeroth mode is varied from device to device.
				Single mode control of the splitting is maintained.
			}
		\end{scope}
	\end{tikzpicture}%
	\caption{%
		\textbf{Meta-dispersion in PhCR}
		\processCaptions
	}
	\label{fig:MetaDisp}
\end{figure*}
\begin{figure}[h]
	\begin{tikzpicture}[inner sep=0, outer sep=0]
		\node[below=1em of current bounding box] (img)  {\includegraphics[width=\linewidth]{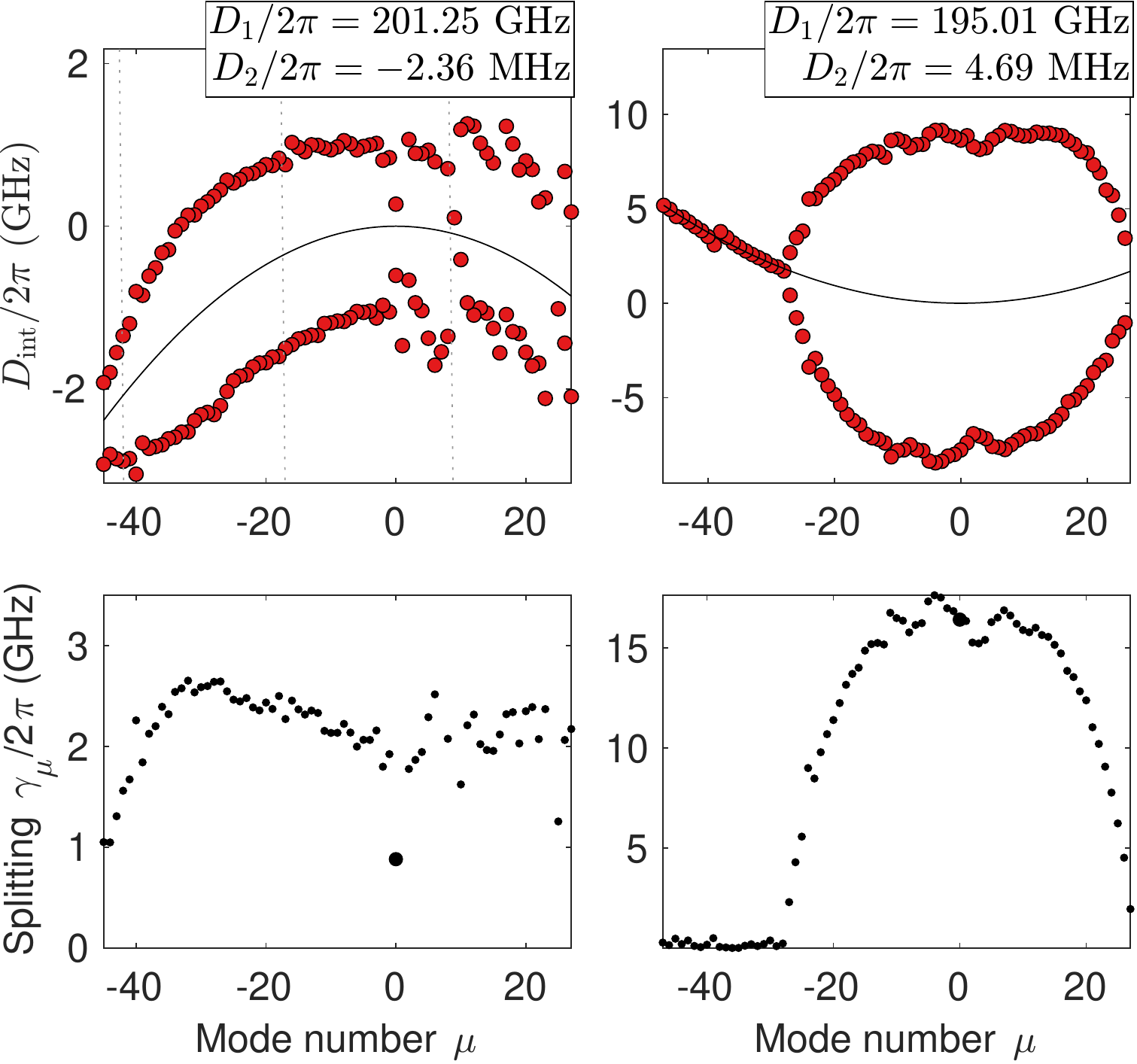}};

		\begin{scope}[shift={(img.south west)},
				x={($0.01*(img.south east)$)},
				y={($0.01*(img.north west)$)}
			]
			\addsubcap{fig:MetaDisp2:DintBlueFlat}{10}{94.5}{%
				Normal dispersion ring with flattening of the blue-shifted (upper) branch.
				The impact of the avoided mode crossings with higher order mode families are causing the disruptions near $ \mu = 10 $.
			}
			\addsubcap{fig:MetaDisp2:gammaBlueFlat}{10}{43}{%
				Corresponding mode splitting distribution.
			}
			\addsubcap{fig:MetaDisp2:DintAnomGaussPulse}{59}{94.5}{
				Anomalous dispersion ring (rw = \SI{1.3}{\micro\meter}) and engineering the red-shifted branch to replicate the profile in \cref{fig:inverselle:Case1D}, and
			}
			\addsubcap{fig:MetaDisp2:gammaAnomGaussPulse}{59}{43}{
				Corresponding mode splitting distribution.
			}
		\end{scope}
	\end{tikzpicture}
	\caption{%
		\textbf{Examples of meta-dispersion in PhCR}
		\processCaptions
	}
	\label{fig:MetaDisp2}
\end{figure}
\begin{figure*}[!tb]
	\centering
	\begin{tikzpicture}
		\node[below=1em of current bounding box,inner sep=0] (img)  {\includegraphics[width=\textwidth]{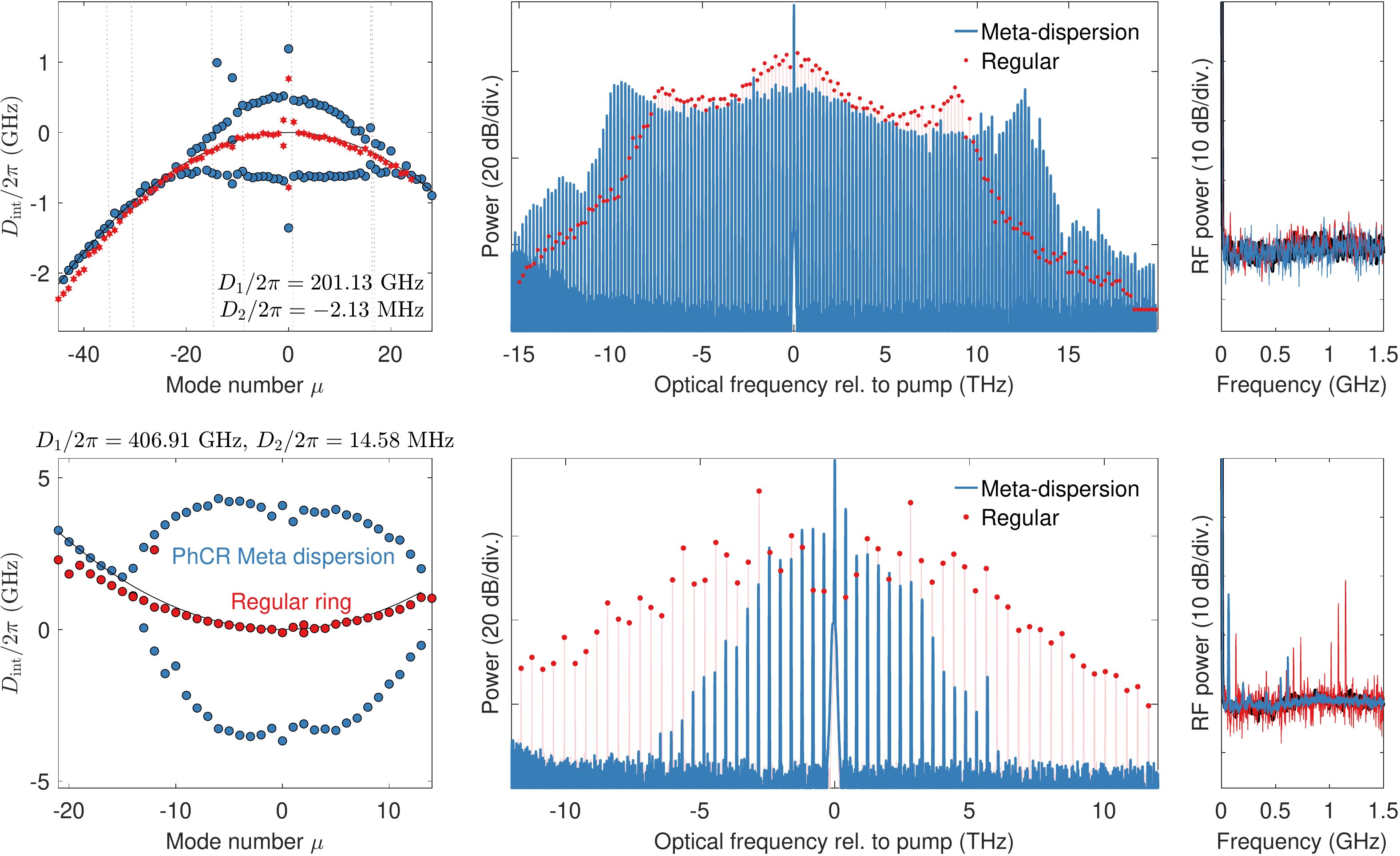}};

		\begin{scope}[shift={(img.south west)},
				x={($0.01*(img.south east)$)},
				y={($0.01*(img.north west)$)},
				inner sep=0, outer sep=0
			]
			\addsubcap{fig:MetaCombs:flatCombDint}{5.1}{98}{
				Measured integrated dispersion.
				The meta-dispersion resonator (in blue) features a flattened and slightly anomalous dispersion near the centre modes (on the down-shifted branch), in an overall normal dispersion ring. A normal dispersion ring with a single mode splitting at $ \mu=0 $ that allows comb generation, is shown in red.
			}
			\addsubcap{fig:MetaCombs:flatComb}{33.5}{98}{
				Optical frequency comb measured on an optical spectrum analyser, obtained in the meta-dispersion resonator (blue) and regular PhCR (red). The on-chip pump power is \SI{55}{\milli\watt} and \SI{56}{\milli\watt} respectively.
			}
			\addsubcap{fig:MetaCombs:flatCombRF}{84.5}{98}{
				The intensity noise of the combs shown in \subref*{fig:MetaCombs:flatComb}~(blue, red) coincides with the electrical spectrum analyzer's noise floor (black), confirming a coherent mode-locked state.
			}
			\addsubcap{fig:MetaCombs:gaussCombDint}{5.1}{44.5}{
				Integrated dispersion of a PhCR microresonator for Gaussian shaped comb generation (blue), compared with a regular ring with anomalous GVD (red).
			}
			\addsubcap{fig:MetaCombs:gaussComb}{33.5}{44.5}{
				Comb states generated in each ring, (PhCr in blue, regular in red) in the noisy modulation instability regime.
				Pump powers are \SI{91}{\milli\watt} and \SI{89}{\milli\watt} respectively.
			}
			\addsubcap{fig:MetaCombs:gaussCombRF}{84.5}{44.5}{
				Intensity noise of the combs shown in \subref*{fig:MetaCombs:gaussComb}.
				The peaks indicate MI instabilities.
			}%
		\end{scope}
	\end{tikzpicture}
	\caption{%
		\textbf{Comb generation in meta-dispersion PhCR}
		\processCaptions
	}
	\label{fig:MetaCombs}
\end{figure*}

\subsubsection{Evolution step}

The GA loop for finding the optimal parameters consists of two steps, presented in \cref{fig:inverselle:algo}.
First, the steady-state intra-resonator field is computed for each member of the population by integrating the LLE equation via split-step Fourier transform.
Next, the steady-state comb solutions are ranked based on an error function $\mathcal{E}(\tilde{A}_\mu)$ that measures their fitness for the optimization goal.
This metric can be the mean square error to a target comb shape or a binary criterion that checks whether a given power target is reached for certain lines.
To ensure stable and low noise combs, highly fluctuating states are penalized.
If a solution meets a fitness threshold or a maximum generation number is reached, the algorithm stops.
Otherwise, a new population is created by combining two parents' dispersion coefficients (crossover) and adding a random variation (mutation) to generate each new candidate, before the loop continues.
The GA is further detailed in the Methods section.

\subsubsection{Optimization results}
First, we show how the dispersion can be optimized to obtain a comb with nearly Gaussian spectral profile in the resonator.
\Cref{fig:inverselle:Case1C} shows the target comb, the initial comb and the result after optimization.
The initial dispersion profile, shown in \cref{fig:inverselle:Case1D}, is calculated from the target comb Kerr shift using \cref{eq:RCC}.
The optimized dispersion appears almost identical to this initial profile. The GA mainly modified the pumping conditions to fit the spectral width to the target. 

For applications, such as telecommunications, a flat microcomb with high conversion efficiency and maximum power per line is typically required.
Using a rectangular comb target is problematic as the function is discontinuous. A raised-cosine shape makes a more realistic target for optimization as it has a smoother decay outside the bandwidth of interest, as shown in \cref{fig:inverselle:Case2C}.
Moreover, the regularity of the function makes it possible to compute the Kerr shift, which again provides an excellent starting point (see \cref{fig:inverselle:Case2D}).
The GA mainly modifies the pump mode shift  $ \gamma_0 $ and the driving parameters.
These examples demonstrate that our method can perform a `fitting' of a specific comb target by changing the parameters of the LLE.
Our initialization technique also proves to be very effective in coarsely matching a desired comb shape, which can speed up the convergence of the GA, especially if it preferentially adjusts the pumping parameters.
An exhaustive search could even be conducted.

An alternative approach to obtain efficient flat comb generation, is to modify the fitness metric to enforce a comb line power of at least $ P_{\rm min} $ over a given bandwidth (see methods).
We initialize the dispersion from the Kerr shift of a flat-top comb model (see the method section), which does not satisfy the power criterion, although we sought to obtain the highest possible power.
After the optimization, the power criterion is met by the resulting comb, as shown in \cref{fig:inverselle:Case3C}.
Interestingly, the associated dispersion profile, shown in \cref{fig:inverselle:Case3D}, is dominantly normal for higher relative mode number, and turns locally anomalous near the pump mode.
The general comb shape resembles the typical `platicon' spectrum obtained in normal dispersion~\cite{Godey2014,Lobanov2015,Yu2021}, i.e. a strong central lobe surrounded by two pronounced wings, which is known to have high conversion efficiency~\cite{Helgason2020}.
However, the center lobe here has a flattened $ \operatorname{sech} $-like profile~\cite{Herr2013}.
In this optimization problem, where the target comb is not provided, the GA plays a prominent role since it extensively explores the parameter space and converge to a radically different solution from the initialization.
Overall, the heuristic approach and the GA are complementary tools in the inverse-design process.
The former can be used to quickly obtain a rough estimate of the optimal parameters in specific problems, while the latter can refine the parameters and explore a broader range of solutions.
Finally, it should be noted that the algorithm does not guarantee to find a global optimum to the problem.
However, we tested the convergence robustness by repeating this optimization problem three times, and obtained similar dispersion profiles for all iterations.

An additional scenario, targeting enhanced dispersive-wave emission is discussed in the supplementary information (SI).
Before finalizing the meta-dispersion design, bidirectional LLE simulations~\cite{Skryabin2020} are performed, with sweeps of $ \delta\omega $, $ P_{\rm in} $ and $ \gamma_0 $ to ensure that the desired nonlinear state can be achieved in the resonator, under realistic driving conditions (see the SI for details).

\subsection{Meta-dispersion}
The complex dispersion obtained with the GA is realized by means of engineered mode splittings in a PhCR (see \cref{fig:concept:PhCR}).
A distributed corrugated grating is fabricated on the inner wall of a ring resonator, introducing a clockwise -- counterclockwise coupling and hence a mode hybridization.
The corrugation's period and amplitude respectively control the impacted longitudinal mode and the splitting amplitude.
By appropriately superposing multiple corrugations, we generalize this concept to simultaneously control multiple dozens of modes, while faithfully retaining a designed spectral distribution of mode splitting $ \gamma_{\mu} $.
The combined ring waveguide dispersion and mode splitting distribution define an effective synthetic dispersion, or \emph{meta-dispersion}, on the blue and red shifted branches corresponding to the up- and down-frequency-shifted split modes.

\subsubsection{PhCR design}
\Cref{fig:MetaDisp:PhcDesign} shows the implementation of the optimization result using meta-dispersion.
The dispersion is decomposed into two components.
First, the ring's dispersion is chosen to match the general sign of the optimized dispersion at high mode numbers, which allows us to set the waveguide's cross-section.
The mode splitting spectral distribution $ \gamma_{\mu} $ near the central modes is then taken as twice the difference between the target dispersion and that of the waveguide, so that one of the shifted branches coincides with the target dispersion.
Finally, the corrugation's azimuthal profile $ \rho(\phi) $ is calculated by use of Fourier synthesis of this distribution:
\begin{equation}
	\rho(\phi) = \sum_{\mu\in \mathbb{Z}} \frac{\tilde{\rho}_{\mu}}{2} \, \cos\left( 2 (m_0 + \mu) \, \phi + \xi \mu^2 \right),
\end{equation}
where the modulation amplitude $ \tilde{\rho}_{\mu} $ is determined based on an experimental calibration of the frequency splitting vs PhC amplitude relation $ \gamma(\rho) $ (see methods). $ m_0 $ is the `carrier' modulation which sets the azimuthal mode index of the central mode $ \mu=0 $.

Importantly, the various frequency components of the grating are superposed with a phase offset (chirp $ \xi $) to distribute the corrugation's amplitude evenly across the ring perimeter.
This minimizes the corrugation peak amplitude and preserves the shape of the desired splitting distribution.
Indeed, since we use the fundamental transverse electric mode of the waveguide, which has the highest Q factor, the relationship $ \gamma(\rho) $ is non-monotonic.
A similar dispersion control approach was developed concurrently in ref.~\cite{MoilleFourierSynthesis2023}, using the transverse magnetic mode, for which $ \gamma(\rho) $ is monotonic.

\subsubsection{Characterization}
We fabricate the designed devices in a 570~nm-thick \ce{Ta2O5} (tantala) layer on thermal silicon oxide, without top-cladding (see~\cite{Jung2020} for fabrication details).
The ring resonator baseline dispersion is controlled by changing the width of the waveguide (rw).
The resonator FSR are selected as 200 or 400~GHz.
\Cref{fig:MetaDisp:SEM} shows a scanning electron microscope (SEM) image of a completed PhCR, revealing the corrugation pattern.

The dispersion and quality factors are characterized by use of a scanning laser spectroscopy method~\cite{Li2012b}, and the resonances frequencies are detected and the integrated dispersion $ D_{\rm int} $ is retrieved.
We first perform a meta-dispersion demonstration with a Gaussian splitting profile in a normal dispersion ring, shown in \cref{fig:MetaDisp:DintGauss}.
The integrated dispersion $ D_{\rm int} $ illustrates the deviation of each frequency splitting branch from the continuous resonator dispersion.
The resonance pair for each longitudinal split mode is fit with a model to recover the characteristics of the resonances.
Thus, we found that adding the corrugation does not degrade the quality factor for modulation depths under \SI{10}{\percent} of the ring width.
The splitting spectral profile $ \gamma_{\mu} $ is also retrieved and shows a Gaussian shape in excellent agreement with the design.

This characterization is repeated for several resonators, in which only the zeroth mode splitting is changed.
Their respective splitting distribution is reported in \cref{fig:MetaDisp:gammaPumpTune}, which shows first the excellent reproduction of the global Gaussian profile across the devices.
Second, we observe that single mode splitting control is accurately retained, in spite of splitting multiple tens of adjacent modes.

\Cref{fig:MetaDisp2} shows further examples of meta-dispersion demonstrations, with various configurations of base ring dispersion (normal and anomalous) and to tailor the dispersion of the red- or blue-shifted branch.
The principle works remarkably well in all cases, such as flattening the blue-shifted branch or carve a localized stronger dispersion.
Deviations from the measured profiles are nevertheless observed, which are caused by avoided mode crossing. These are due to the presence of higher order modes in the resonator and can be problematic for the formation of DKS~\cite{Herr2013b}.

\subsection{Comb generation}
We conducted comb generation experiments in PhCR structures with meta-dispersion profiles for two configurations of target comb spectra.
To initiate comb generation, the laser power is increased, and its frequency is swept through the red-shifted resonance of the split mode $ \mu=0 $, starting from the center of the split and increasing the laser wavelength.

The first configuration, targets a comb with a minimum comb line power.
The measured dispersion, shown in \cref{fig:MetaCombs:flatCombDint}, corresponds to that of a normal-dispersion ring resonator ($ \text{rw}= \SI{2.3}{\micro\meter} $, radius \SI{109}{\micro\meter}) with a slight anomalous meta-dispersion near the pump, to reproduce the optimized design shown in  \cref{fig:inverselle:Case3D}.
The device was pumped with \SI{55}{\milli\watt} on-chip.
The detuning was first increased to trigger the formation of an initial comb state  and then reduced (backward tuning), to induce switching into the state shown in \cref{fig:MetaCombs:flatComb}.
The spectrum is analogous to a platicon comb~\cite{Lobanov2015}, but with a flattened center lobe, compared to a single PhC device (shown in red in \cref{fig:MetaCombs:flatComb}), and a wider bandwidth for the same pump power and base ring dispersion.
The number of consecutive teeth within the 5-dB bandwidth is increased from $ \sim 20 $ to $ \sim 50 $.
Notably, the power per line in the central lobe is lower than in the wings, as featured in the predicted inverse-design comb in \cref{fig:inverselle:Case3C}.
We confirmed that this comb is a low-noise state, by recording its intensity noise, which coincides with the instrument's noise floor (\cref{fig:MetaCombs:flatCombRF}).
Several comb state transitions and breathing stages were observed by changing the power and detuning (see SI), revealing complex dynamics.

The second configuration aims at generating Gaussian-shaped combs.
The ring features anomalous base dispersion ($ \text{rw}= \SI{1.55}{\micro\meter} $, radius \SI{53}{\micro\meter}), and the meta-dispersion is chosen to replicate the computed optimized dispersion in \cref{fig:inverselle:Case1D} over a limited range of the red-shifted branch, as shown in \cref{fig:MetaCombs:gaussCombDint}.
This meta-dispersion PhCR device is compared to a regular ring resonator with the same geometry to provide a baseline for comparison.
\Cref{fig:MetaCombs:gaussComb} shows the combs produced in each resonator in the unstable modulation instability (MI) regime, when pumping with \SI{\sim 90}{\milli\watt} on-chip.
We also ensured that the comb states have comparable intensity noise properties and are in the same stage of MI (subcomb merging~\cite{Herr2012}).
The regular ring produces a typical MI comb, with powerful comb lines surrounding the pump (corresponding to primary lines) and a dip in comb line power near the pump.
The meta-dispersion resonator exhibits a markedly different comb envelope, with a smoother decrease in power from pump to wings.
The MI comb bandwidth is also largely reduced, highlighting the impact of the PhCR in shaping the microcomb.
Unfortunately, stable mode-locked soliton sates~\cite{Herr2012} could not be obtained in this meta-dispersion device, nor in the regular ring, likely due to the photothermal effect in the cavity~\cite{Li2017a,GaoProbingMaterial2022}, which is particularly strong in \ce{Ta2O5}.
The presence of an avoided mode crossing near the pumped mode can also hinder soliton formation~\cite{Herr2013b}.
Therefore, the PhCR comb envelope does not match the expected Gaussian shape, which is obtained in the soliton state, as shown below.
Nevertheless, we believe that these limitations are not fundamental, but of technical nature, and that the desired states can be achieved on a more mature technology platform.

\subsubsection{Comb simulation}
We have compared these experimental observations with bidirectional LLE simulations~\cite{Skryabin2020}, detailed in the SI.

First, we set out to reproduce the comb states in noisy MI, in order to validate our model.
The parameters used are taken from the experimental measurements of the resonators shown in \cref{fig:MetaCombs:gaussCombDint}.
The detuning is the only adjusted parameter, to recover the unstable modulation instability regime in each resonator.
The averaged spectra shown in \cref{fig:MetaCombsSims:MI} are in good agreement with our experimental observations.

\begin{figure}[ht]
	\centering
	\begin{tikzpicture}[inner sep=0, outer sep=0]
		\node[below=1em of current bounding box] (img)  {\includegraphics[width=.9\linewidth]{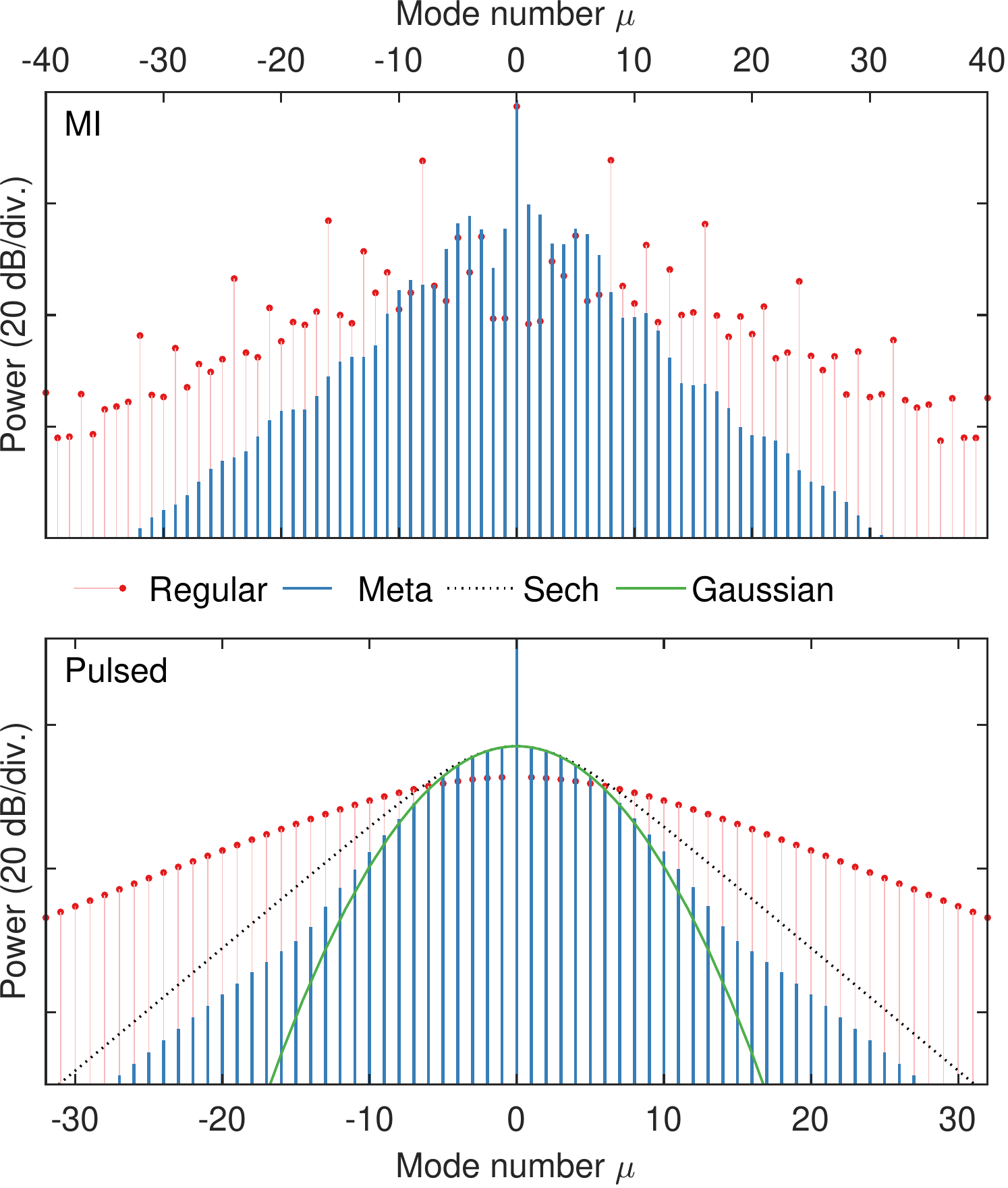}};
		
		\begin{scope}[shift={(img.south west)},
			x={($0.01*(img.south east)$)},
			y={($0.01*(img.north west)$)}
			]
			\addsubcap*{fig:MetaCombsSims:MI}{-2}{92}{
				Simulated microcombs in the noisy MI regime, for the PhCR (blue) and regular anomalous GVD ring (red) presented in \cref{fig:MetaCombs:gaussCombDint}.
				The spectra are averaged over \num{\sim 130000} roundtrips.
			} 			%
			\addsubcap*{fig:MetaCombsSims:Pulsed}{-2}{47}{
				Simulation of microcombs in the low-noise pulsed (soliton) regime, for the PhCR (blue) and regular anomalous GVD ring (red).
				Over the meta-dispersion bandwidth, the PhCR comb features a Gaussian shape (green), as designed.
			}
		\end{scope}
	\end{tikzpicture}
	\caption{%
		\textbf{Numerical simulation of comb states}
		\processCaptions
	}
	\label{fig:MetaCombsSims}
\end{figure}

Second, since numerical simulations are not subject to thermal effects, they allow us to extrapolate the shape of the stationary `mode locked' pulsed states in each resonator, which are reached at higher detuning (see SI).
The corresponding comb spectra are overlaid in \cref{fig:MetaCombsSims:Pulsed}.
The PhCR comb clearly deviates from the usual DKS $ \operatorname{sech}^2 $ profile formed in the standard resonator, and follows a Gaussian shape within the meta-dispersion bandwidth.
These projections further show that the nonlinear cavity field state is governed by the meta-dispersion.

The results presented here show only the combs co-propagating with the pump.
Since the combs are supported by hybrid modes, whose splitting amplitude is significantly larger than the Kerr shift~\cite{Yu2020,Yu2021}, they are typically generated with near equal power in the clockwise and counterclockwise direction.
However, transitions to a dominant direction have been observed, especially when the dispersion is predominantly normal.

\section*{Discussion}
In summary, we have demonstrated the spectral shaping of microcombs by inverse design, using an optimization layer on the LLE, and we have demonstrated the scalable use of PhCRs for effective meta-dispersion engineering.
This flexible approach enables mode-by-mode selectivity and control of nonlinear frequency shifts that are critical in Kerr microcombs.
We also showed that meta-dispersion can be used in the nonlinear regime to perform spectral shaping of microcombs.

Further fundamental studies are still needed to fully understand the nonlinear dynamics in hybrid modes, to help control comb formation in these new devices, and to refine our inverse design approach.
Moreover, we are considering several ways to enhance our method.
The optimization algorithm could occur directly on the meta-dispersion parameters, using the bidirectional LLE within the evolution loop.
This requires nonetheless a robust method of seeding the cavity field in both directions.
In this respect, machine learning could contribute to speed up the computation, and robustly find the soliton or related nonlinear eigenstates.
Other implementations of the optimised dispersion can be conceived, based on coupled resonator chains~\cite{JiEngineeredZerodispersion2023,Helgason2020}, through engineered avoided mode crossing with higher-order modes~\cite{Wang2020b}, or by using discrete inverse-designed reflectors~\cite{AhnPhotonicInverse2022}.

We believe that our comb customization procedure is promising to fully unlock the application potential of microcombs.
First, in optical telecommunications, where it can maximize the conversion efficiency and the power per line in the bandwidth of interest.
In spectroscopy, our concept can help open access to new wavelength ranges, or shape combs that target specific chemical species.
Additionally, one could try to optimize the control of the comb parameters (repetition frequency, offset), to improve its intrinsic stability.
Finally, the presented method offers new avenues for nonlinear physics, making previously unrealistic dispersion profiles accessible, while contra-propagative coupling promises to produce a myriad of complex dynamics that remain to be explored.

\section*{Acknowledgments}
E.L. acknowledges support from the Swiss National Science Foundation (SNSF) under contract No.~191705. This research was funded by the DARPA PIPES program under HR0011-19-2-0016, the AFOSR FA9550-20-1-0004 Project Number 19RT1019, and NIST.

\section*{Authors contributions}
E.L. implemented the optimization algorithm, performed the experiments and analysed the data. E.L and S.-P.Y. contributed to the numerical simulations and resonators design.  T.C.B. and D.R.C fabricated the microresonators.
E.L. wrote the manuscript, with input from all authors. S.P. supervised the project.

\section*{Competing Interests}
David R. Carlson is a cofounder of Octave Photonics. The
remaining authors do not currently have a financial interest in \ce{Ta2O5}-integrated photonics.

\clearpage

\appendix
\section*{Methods}
\setcounter{subsection}{0}
\setcounter{figure}{0}
\renewcommand\thefigure{S\arabic{figure}}
\makeatletter
\renewcommand{\fnum@figure}{\textbf{Extended Data Figure~\thefigure}}
\makeatother
\subsection{Genetic algorithm}
Our goal is to find the parameters $ D_k, \gamma_0, \delta\omega, P_{\rm in} $ that yield a stationary comb profile (for the LLE) $ \tilde{A}_\mu $ that minimizes an error function $ \mathcal{E}(\tilde{A}_\mu ) $, which quantifies our objective.
The optimization problem can be formalized as follows:
\begin{equation}\label{eq:optimProblem}
	\left[
	\begin{aligned}
		& \underset{D_k, \gamma_0, \delta\omega, P_{\rm in}}    %
		{\operatorname{minimize}} \; \mathcal{E}(\tilde{A}_\mu ) \\
		& \text{subject to} \;
		\left\lbrace
		\begin{aligned}
			& \text{\cref{eq:LLEeq}} = 0            \\
			& P_{\rm in} \leq P_{\rm max}
		\end{aligned}
		\right.
	\end{aligned}
	\right.
\end{equation}
The latter term contrains the pump power within a limited power budget.
Based on the available power and our resonator's properties, we set $ P_{\max} = 13 \, P_{\rm th} $.

The LLE steady state field for each member of the population, is estimated using split step Fourier transform integration~\cite{Hansson2014} of \cref{eq:LLEeq}.
An initial pulse consisting in an approximate soliton profile~\cite{Coen2013}, is seeded in the simulation and its evolution through the LLE is computed for 100 photon lifetimes; the pump power and detuning are kept constant.
Though the stationary state of the LLE can be found using a Newton--Raphson method~\cite{Coen2013}, we found that its convergence was slower and less systematic than the split step approach, especially in complex dispersion profiles.

The steady-state solution of each member of the population is then ranked according to a metric $ \mathcal{E}(\tilde{A}, D_k) $, which measures each comb's `fitness' with respect to the optimization goal.
The smaller the error, the more fit the candidate.
This metric can be the squared Euclidean distance to a target comb shape $ \tilde{A}_\mu^{\rm target} $:
\begin{equation}\label{MSE}
	\mathcal{E}(\tilde{A}) = \sum_{\mu\neq0} \Bigl( |\tilde{A}_\mu| -  |\tilde{A}_\mu^{\rm target}|  \Bigr)^2
\end{equation}%
This metric can be complemented to favour stable low-noise states, by adding the standard deviation of the mean comb power, estimated during the second half of the split-step evolution. 

The raised cosine comb spectrum shape is defined with the following expression:
\begin{equation}\label{eq:RCC}
	\tilde{A}_\mu^{\rm target} =
	\begin{cases}
		1,
		& \frac{|\mu|}{\Delta} \leq (1-\beta)
		\\
		\cos^2\left[{\frac {\pi}{4 \beta} \left(\frac{|\mu|}{\Delta}-{(1-\beta)}\right)}\right],
		& \abs{\frac{|\mu|}{\Delta} -1 } \leq \beta
		\\0,
		& {\text{otherwise}}
	\end{cases}
\end{equation}
where $ \Delta $ is the comb's 3-dB bandwidth and $ \beta $ is the roll-off factor, which controls the slope of the power decay in the outer lines of the comb.

Another metric can be used to ensure a minimum comb line power $ P_{\rm min} $ over a given bandwidth $\Delta$:
\begin{equation}\label{minPowMetric}
	\mathcal{E}(\tilde{A}) = \sum_{\mu \in \Delta}\max\left (0, \,  P_{\rm min}  -  \dfrac{\hbar\omega_\mu \, D_1}{2\pi} \, |\tilde{A}_\mu|^2\right)
\end{equation}
This way, comb lines exceeding the criterion are not penalized.
A soft threshold function can also be used to force the power to be within a given range.
The flat-top comb initial model follows the equation $ A_\mu = A_0 / (1+\mu/B)^8 $, where $ B $ sets the bandwidth and $ A_0 $ the comb line magnitude. $ A_0 $ was set to the maximum value authorizing convergence to a stable comb state.

In general, engineering the error function is a key aspect of inverse design problems, and it requires careful consideration of the optimization goals, constraints, and trade-offs.
The two functions in \cref{MSE,minPowMetric} are two extreme cases when is come to evaluating the combs shape over a bandwidth of interest, as the former gives equal weights to all comb lines, while the latter only accounts for the lines within a spectral region and disregards the lines outside.
A more general  approach is to introduce weight factors that reflect the importance of certain comb lines or regions of the spectrum.
Alternatively, one could also use a weighting function that assigns higher importance to certain dynamics of the comb, to identify phenomena such as breathing.

Note that if the targeted problem is frequency-symmetric with respect to the pump, all odd dispersion coefficients can be set to zero and the optimization can be performed only on the even terms, reducing the number of variables while ensuring the symmetry of the solution.

At each iteration of the GA loop, a new population is created.
First the two best individuals from the last generation are carried over unchanged (elitism), which guarantees that the current optimum is preserved.
Then, each new offspring is generated by selecting two (or more) parents at random, with a probability inversely proportional to their residual error, so that the fittest to the problem are selected more frequently (roulette wheel selection).
The dispersion coefficients (genes) of the offspring are randomly selected from its parents (uniform crossover).
The better the fitness of a parent, the more likely its genes will be selected.
Finally, the offspring's genes are mutated by adding a random value drawn using a normal distribution with standard deviation $ \sigma $.
As the number of generation increases, $ \sigma $ is gradually decreased to favour the convergence of the algorithm.

Our genetic evolutionary algorithm was developed in MATLAB, without the use of existing optimization libraries. To speed up the steady-state intra-resonator field calculation, we implemented parallelization of the split-step Fourier transform computations. This allowed us to significantly reduce the computation time, making the optimization process feasible within a few hours.

\subsection{Mode splitting calibration}
\begin{figure}[htb]
	\centering
	\includegraphics[width=0.9\linewidth]{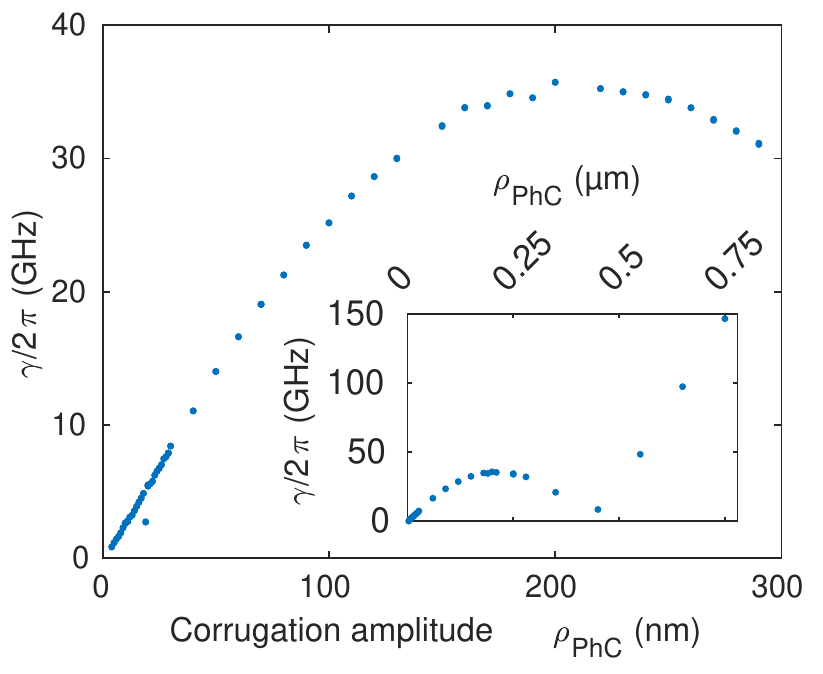}
	\caption{
		\textbf{PhC and mode splitting calibration}.
		Measured mode splitting of the target mode as a function of corrugation amplitude $\rho_{\rm PhC}$ (single frequency PhC), measured across 114 resonators.
		The inset shown an extended range of $\rho_{\rm PhC}$.
	}
	\label{fig:SI:gamma_APhC}
\end{figure}

The relation between PhC amplitude and mode splitting was calibrated experimentally on a series of single period PhCR with identical base geometry (ring radius \SI{109.5}{\micro\meter}, width \SI{2}{\micro\meter}, height \SI{570}{\nano\meter}), while the PhC amplitude $ \rho_{\rm PhC} $ is swept.
The results are shown in \cref{fig:SI:gamma_APhC}. Note that the reported amplitude corresponds to the design value and not the measured geometry.

While the relation is linear at small corrugation amplitudes, it reaches a maximum around $ \rho_{\rm PhC} \sim\SI{200}{\nano\meter} $ for $ \gamma/2\pi \sim \SI{35}{\GHz}$ and the tuning direction reverses.
We found that a near null splitting can be obtained for $ \rho_{\rm PhC} \sim\SI{420}{\nano\meter} $, before $ \gamma $ increases again for larger amplitude values (see inset of \cref{fig:SI:gamma_APhC}).
This counter-intuitive bandgap closing was observed and explained by Gnan et al. in photonic wire Bragg gratings and is related to Brewster angle incidence~\cite{Gnan2009} and is also studied in ref.~\cite{MoilleFourierSynthesis2023}.

\begin{figure}[htb]
	\centering%
	\includegraphics[width=\linewidth]{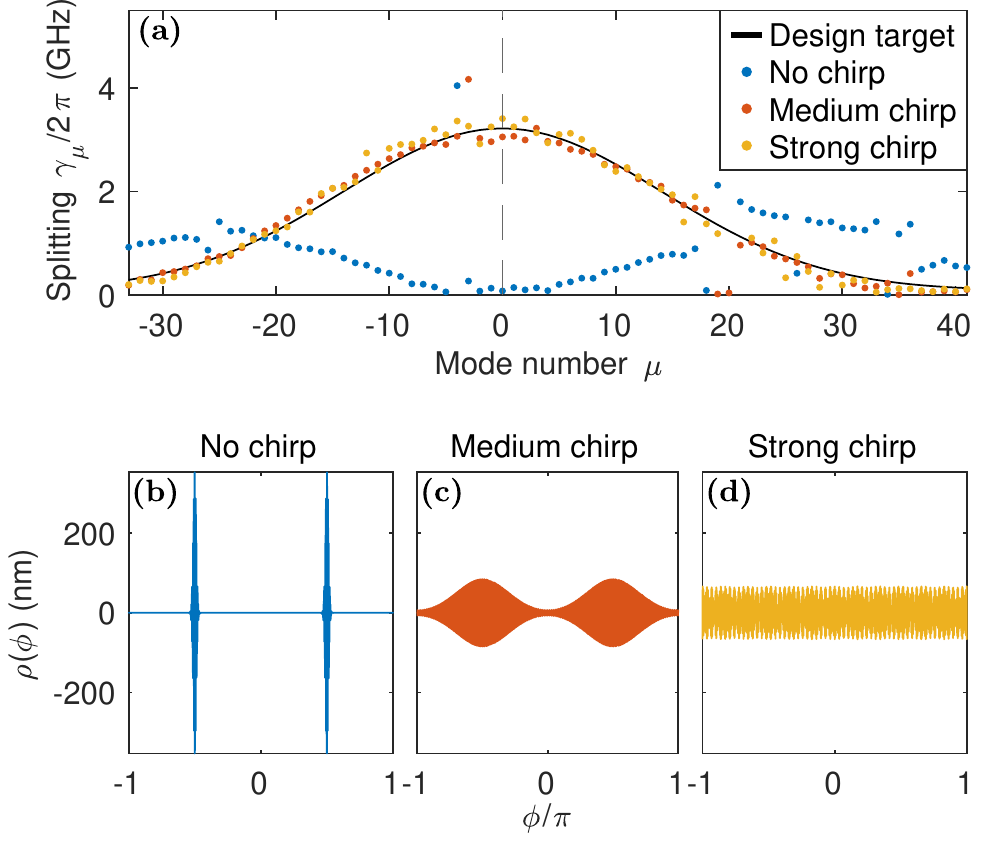}%
	\caption{
		\textbf{Effect of distributing the PhC pattern along the ring perimeter}.
		\textbf{(a)} Measured mode splitting distribution with various amount of chirping applied to the corrugation pattern. The design target is the Gaussian distribution shown in black. \textbf{(b-c)} Spatial profiles of the designed corrugations. \textbf{(a)} Without chirp, the splitting distribution is distorted and even inverted. The effect is avoided by chirping the pattern.
	}
	\label{fig:SI:chirpImpact}
\end{figure}

We did not observe any significant deviation from this calibration, upon splitting multiple modes.
Chirping the frequency components of the corrugation, in order to reduce amplitude variations, allows us to stay within the linear segment of this relationship and ensures a good replication of the splitting profile, as shown in \cref{fig:SI:chirpImpact}.
Limiting the amplitude variations in the corrugation also avoids the need for high aspect ratio etching, which thus simplifies the nanofabrication of the PhCR.

\section*{Data availability statement}
The data and code used to produce the figures of this manuscript are available on Zenodo: \url{https://doi.org/10.5281/zenodo.7998103}.

\section*{Code availability statement}
The code for the genetic algorithm implementation used to perform the dispersion optimization is available at: \url{https://github.com/ErwanLucas/inverseLLE}.

\vfill\null
\clearpage



\section*{Supplementary material}

# Supplementary information to: Tailoring microcombs with inverse-designed, meta-dispersion microresonators

Erwan Lucas*[1,2,4], Su-Peng Yu[1,2], Travis C. Briles[1], David R. Carlson[1,3], and Scott B. Papp†[1,2]

[1] *Time and Frequency Division, National Institute of Standards and Technology, Boulder, CO USA*
[2] *Department of Physics, University of Colorado, Boulder, CO 80309, USA*
[3] *Octave Photonics, Louisville, CO 80027, USA*
[4] *Current affiliation: Laboratoire ICB, UMR CNRS 6303, 21078 Dijon, France*

## Contents

## Octave-spaced dispersive wave optimization

We consider the case of optimizing the dispersion and pump driving parameters to maximize the power at a frequency located an octave above the pump, similarly to the scenario presented in ref. [1].

This configuration can facilitate Kerr comb self-referencing, as the high power pump can be more easily frequency doubled by second harmonic generation and compared to the comb line located one octave away. An elegant way to enhance the comb line power is to use dispersion engineering to create a phase-matched condition near the desired frequency to trigger the emission of a dispersive wave (DW) by a DKS [2]. However, finding the appropriate dispersion and driving conditions to optimize the power at a specific wavelength target is not an easy task, and we use our optimization algorithm to fine tune these conditions. Note that similar approaches were applied to supercontinuum generation [3, 4].

We assume here a comb line spacing of 200 GHz, pumped at 1550 nm, leading to an octave relation of the pump with the mode $\mu = 969$, and use 4 orders of dispersion. We limit the pump power to $P_{\text{in}}/P_{\text{thresh}} \leq 50$. The initial dispersion is set by placing a root of $D_{\text{int}}$ at the desired mode position, which in first order, should yield a phase matching condition at that point [5]. We can see in fig. 1 that this initial dispersion yields a wrong DW position, since the complex non-linear phase shift is not taken into account. Our optimization metric follows eq. (6) in the main text, but a term based on the standard deviation of the temporal waveform was added. It favors pulsed localized states over extended chaotic patterns in the cavity. After optimization, the power at one octave is enhanced by $\sim 49$ dB compared to the initial guess.

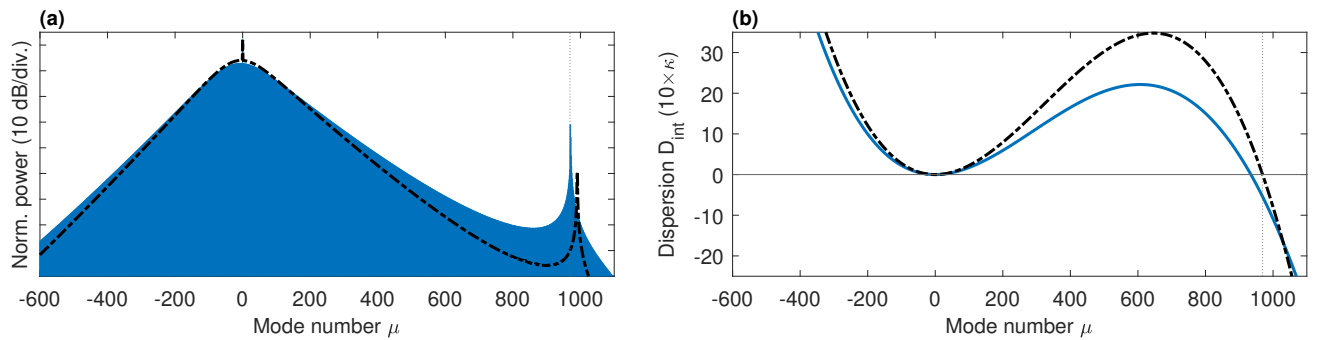

**Figure 1**: **Dispersive wave power optimization**. **(a)** Optimization of the power in a dispersive wave at an octave above the pump (vertical dotted line on mode $\mu = 969$), the dashed black line shows the initial guess. The blue comb is resulting from the optimization. The normalized pump power is $P_{\text{in}}/P_{\text{thresh}} = 43.1$ and detuning $2\Delta\omega/\kappa = 43.8$ **(b)** Initial (black, dashed) and optimized (blue) dispersion curves.

This optimization approach can be generalized and refined. The dispersion can be derived from real waveguides, instead of using polynomials [1]. Moreover, if the Raman self frequency shift [6, 7] of the underlying material is taken into account, the dispersion optimization can adapt accordingly.

---

*erwan.lucas@polytechnique.org
†scott.papp@nist.gov

## Bidirectional, coupled Lugiato-Lefever simulations

We developed numerical simulations of a bidirectional LLE model, based on ref [8]:

$$\begin{cases} \dfrac{d\tilde{A}_\mu^+}{dt} = -\left(\dfrac{\kappa}{2} + i\,\delta\omega - i\,D_2\mu^2\right)\tilde{A}_\mu^+ + i\,g_0\,\mathcal{F}\left(|A^+|^2 A^+\right)_\mu + i\,\dfrac{\gamma_\mu}{2}\tilde{A}_\mu^- + 2i\,g_0\left(\sum_{\mu'}|\tilde{A}_{\mu'}^-|^2\right)\tilde{A}_\mu^+ + \sqrt{\kappa_{\text{ex}}\dfrac{P_{\text{in}}}{\hbar\omega_0}}\delta_{\mu=0} \\ \dfrac{d\tilde{A}_\mu^-}{dt} = -\left(\dfrac{\kappa}{2} + i\,\delta\omega - i\,D_2\mu^2\right)\tilde{A}_\mu^- + i\,g_0\,\mathcal{F}\left(|A^-|^2 A^-\right)_\mu + i\,\dfrac{\gamma_\mu}{2}\tilde{A}_\mu^+ + 2i\,g_0\left(\sum_{\mu'}|\tilde{A}_{\mu'}^+|^2\right)\tilde{A}_\mu^- \end{cases} \quad (1)$$

where the terms are introduced in the main text. The $+/-$ superscript refers to the fields in the clockwise / counter-clockwise directions. The numerical simulations of this system are based on split-step Fourier method.

## Fine tuning the mode splitting spectral distribution

The above model is used to finalize the inverse design stage. After the genetic algorithm has converged, and the optimal dispersion profile has been decomposed into the baseline dispersion of the waveguide and a spectral splitting distribution $\gamma_\mu$, we map the space of achievable solutions using our numerical model, as shown in fig. 2. To that end, multiple detuning scans across resonance are simulated under varying pump power conditions, while the cavity is only seeded with vacuum noise. This 'soft excitation' scheme is close to the experimental generation method, but thermal effects are excluded.

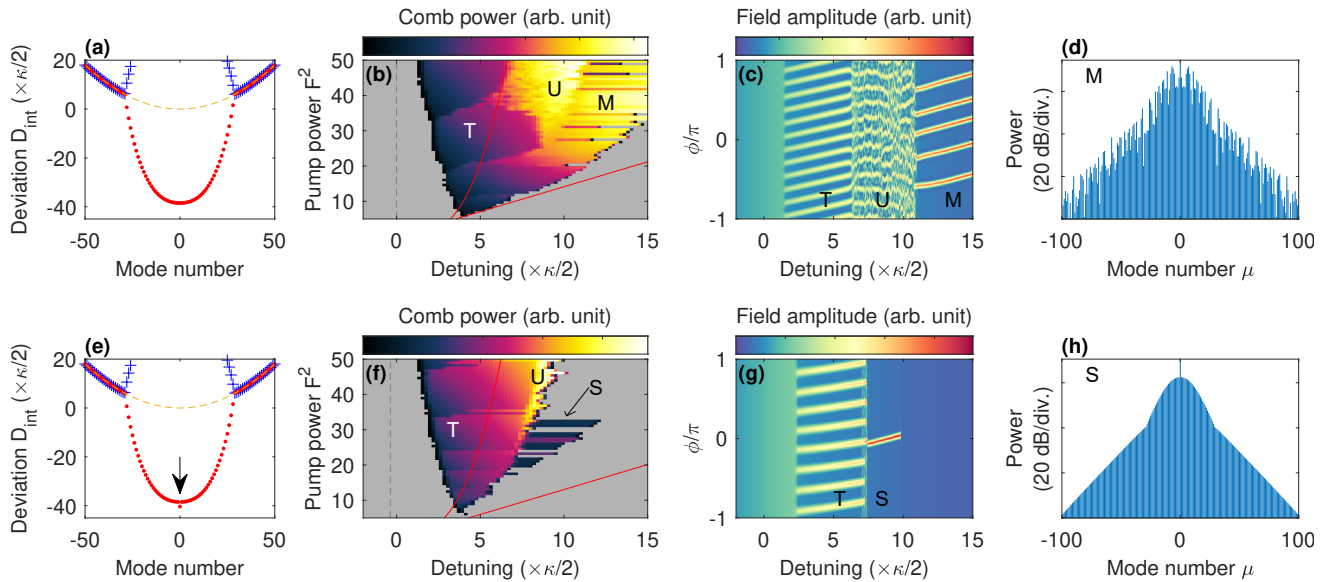

**Figure 2**: **Additional pump shift design**. (a) Meta-dispersion for the Gaussian-shaped comb, without increased splitting of the pumped mode $\gamma_0$. (b) Detuning – normalized pump power phase diagram, showing the normalized comb power. The regions with Turing rolls (T), unstable modulation instability (U) and multi-pulse formation (M) are marked. Only the forward field (+) is shown throughout the figure. (c) Intracavity field as detuning is increased for a normalized pump $F^2 = 45$ showing the transition through the various states. (d) Individual spectrum in the M state shown in (c). (e) Meta-dispersion for the Gaussian-shaped comb, where $\gamma_0$ is specifically increased by $1.87\,\kappa$ (arrow), and (f) corresponding phase diagram, showing "steps" features where a single pulse is formed (S). (g) Intracavity field as detuning is increased for a normalized pump $F^2 = 25$, showing the transition from Turing rolls into a single pulse state. (h) Individual spectrum in the S state shown in the previous inset.

In our optimization algorithm, the field is initialized from a pulse, which does not guarantee that the resulting steady-state localized pattern can be obtained through soft excitation. However, shifting the pump mode can help displace the modulation instability (MI) region [9] to the point where it overlaps with the region of existence of the targeted localized pattern, which can nucleate spontaneously during the soft-excitation.

Thus, the mapping against detuning and power is repeated for several small variations of the pumped mode splitting $\gamma_0$, as illustrated in figs. 2a and 2e. We found that increasing $\gamma_0$ relative to the adjacent modes can help ensure single pulse state generation, as was observed for dissipative Kerr solitons [10]. Figure 2a shows that a small increase of $\gamma_0$ leads to a significant change in the phase diagram of the solution and promotes direct

transitions to the single pulse state at lower power. Since $\gamma_0$ is the only parameter that is changed, an exhaustive simulation is rather fast. This mapping of the parameter space allows us to narrow down the parameter variations needed in the PhCR design. Note that when the dispersion is globally normal, this additional shift gets necessary to establish a comb [11].

## Unusual comb states and dynamics in meta-dispersion PhCR

We provide here additional evidence for the singular comb dynamics in the resonator presented in fig. 5d of the main text. Figure 3 shows several comb states recorded experimentally for different values of detuning and pump power. The observed evolution does not correspond to a classical platicon. In the following, we describe the evolution of several states, labelled in Figure 3a, as the detuning is decreased.

State (i): This initial comb state state appears after tuning the laser across the red-shifted mode for soft excitation. It is broadband but appears to be noisy and breathing, as evidenced by the intensity noise measurement on the RF spectrum analyser (see fig. 3b).

State (ii): When the detuning is decreased, the comb shape changes drastically and takes a slanted flat-top profile. This spectrum shape approaches those obtained in "photonic molecules" [12]. However, the comb state is unstable (likely breathing instability) at low power, but gains stability at higher power (v-vi), which is in contradiction with the usual behaviours of Kerr combs that tend to be destabilized with stronger driving [13].

State (iii): As the detuning is further reduced, the comb switches to a spectrally modulated envelope, which is reminiscent of the spectra obtained in the case of normal, but near-zero GVD [14], but over a smaller spectral width in our case. This behaviour is not entirely surprising, since within the meta-dispersion region the effective dispersion is nearly zero, as can be seem in fig. 3c.

State (iv): Finally, at the minimal detuning the spectrum assumes this platicon-like shape, which was shown in the main text. Strikingly, the most resembling platicon state in a usual normal-dispersion resonator would be expected for a maximal detuning [11].

At present, a complete understanding of the dynamics of combs under meta-dispersion is lacking, and a full study dedicated to these mechanisms is still needed. Nonetheless, we performed bidirectional simulations of the experimentally-measured parameters, which allowed us to qualitatively reproduce some of the observed behaviour, especially for the states (i), (iii) and (iv), as shown in figs. 3d to 3f.

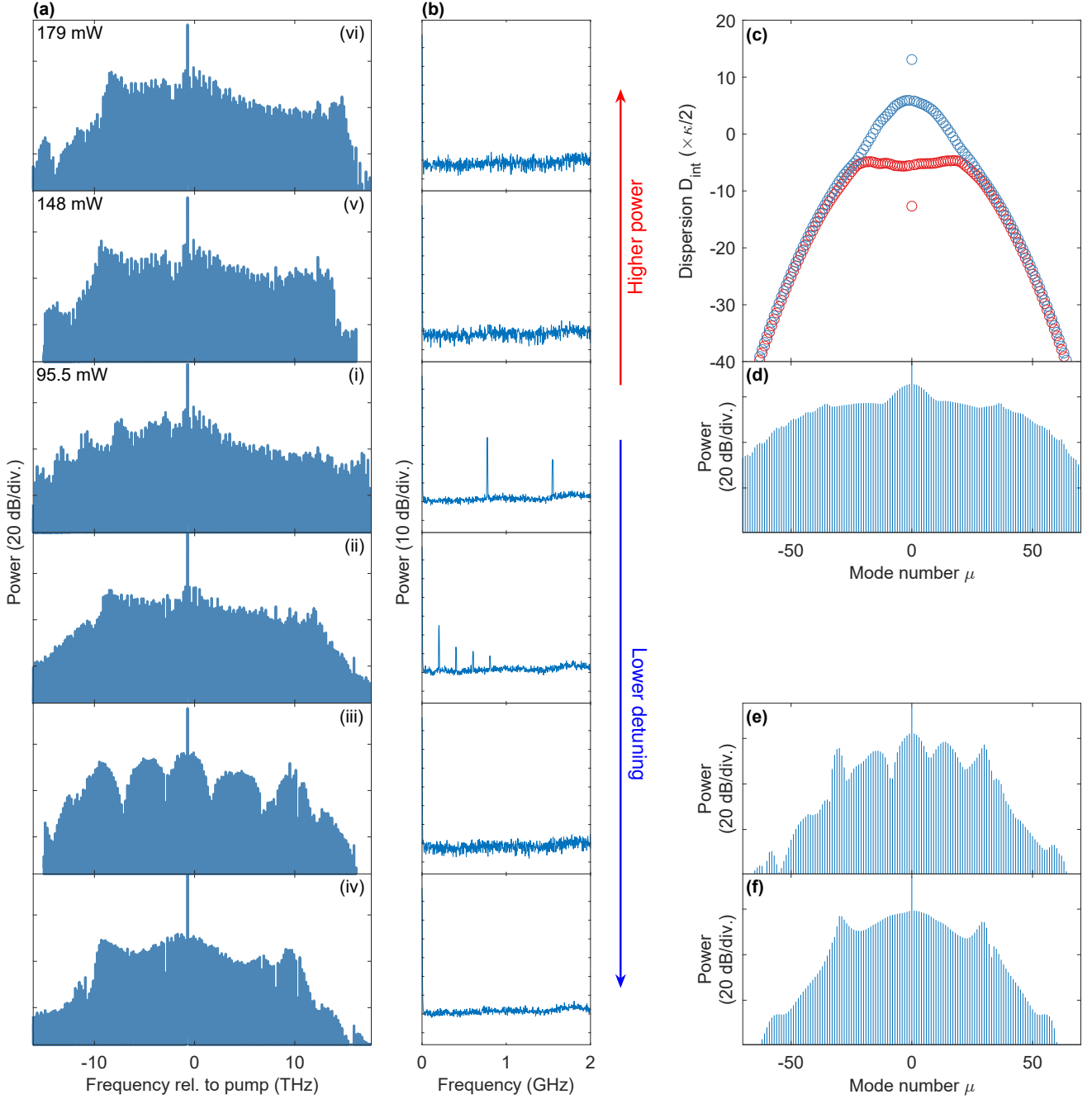

**Figure 3**: **Extended comb characterization in meta-dispersion resonator**. **(a)** Experimentally-measured combs in different conditions of detuning and pump power in the resonator presented in fig. 5d of the main text. The state (i) is the first comb state produced upon soft excitation (pump power 95.5 mW on chip). (i-iv) show the comb evolution while reducing the detuning for the same pump power. The state (iv) is shown in the main text, but at lower power. The states (v-vi) are obtained upon increasing the pump power and for a detuning similar to that of state (ii). **(b)** Measured intensity noise of the combs shown in (a). **(c)** Meta-dispersion profile used for comb simulation, which was retrieved from the measured dispersion. **(d)** Simulated comb for driving parameters $\{2\delta\omega/\kappa, P_{\rm in}/P_{\rm th}\} = \{3.45, 17\}$, resembling the comb state (i) **(e)** Simulated comb for lower detuning $\{-1, 17\}$, which reproduces the modulated spectrum of (iii) **(f)** Simulated comb for lowest detuning $\{-1.4, 17\}$, whose shape matches the state (iv).

4



\end{document}